\documentclass{article}
\usepackage[utf8]{inputenc}
\usepackage{graphicx}
\usepackage{biblatex}
\usepackage{authblk}
\title{Exploring the Efficacy of Transfer Learning in Mining Image-Based Software Artifacts}

\author[1]{Natalie Best}
\author[1,2]{Jordan Ott}
\author[1]{Erik Linstead}

\affil[1]{Machine Learning and Assistive Technology Lab\newline Chapman University}
\affil[2]{Bren School of Information and Computer Sciences\newline University of California, Irvine}
\date{}

\bibliography{sources}

\begin{document}

\maketitle

\begin{abstract} 
    Transfer learning allows us to train deep architectures requiring a large number of learned parameters, even if the amount of available data is limited, by leveraging existing models previously trained for another task. Here we explore the applicability of transfer learning utilizing models pre-trained on non-software engineering data applied to the problem of classifying software UML diagrams. Our experimental results show training reacts positively to transfer learning as related to sample size, even though the pre-trained model was not exposed to training instances from the software domain. We contrast the transferred network with other networks to show its advantage on different sized training sets, which indicates that transfer learning is equally effective to custom deep architectures when large amounts of training data is not available.
\end{abstract}

\section{Introduction}
Despite the recent successes of deep architectures, such as convolutional neural networks, on software engineering data, the lack of sufficiently large training sets for some applications continues to be a substantial hurdle. This requirement has led researchers to label tens of thousands \cite{Ott2018} and even millions of images \cite{russakovsky2015imagenet} by hand. Recent work has shown that this precludes the use of many off-the-shelf convolutional neural network architectures, requiring empirical software engineering researchers to rely on custom (more compact) architectures \cite{Ott2019}. Another possible solution, however, is to leverage transfer learning to deal with large parameter spaces. Through this process models learn in one domain - where data is plentiful - and \textit{transfer} this knowledge to a domain where data is scarce. 

One significant limitation in deep learning is data dependence. As computational ability and available algorithms have improved significantly over the years, many deep learning techniques are still held back by the need for massive amounts of labeled truth data. As architectures increase in depth and number of parameters, the amount of data needed to train networks increases as well. When large datasets are not available, or are difficult to curate, researchers must turn to other methods in order to improve their models. Other possible solutions to small amounts of data have been investigated including low shot learning, meta-learning, and data augmentation \cite{Ott2019}. Although, even with these other methods, the bottleneck of computation time and large parameter spaces remain. 

In this paper, we explore transfer learning as a way to combat the issues related to limited data. Many publicly-available, state-of-the-art models already exist and have been trained on huge amounts of data including VGG \cite{VGG}, AlexNet \cite{AlexNet}, ResNet \cite{ResNet}, and Inception\cite{Inception}. These networks have repeatedly been applied to different tasks from which they were originally trained \cite{Ott2018,shin,Bayramoglu, ott2018learning, alahmadi2018accurately}. We will also apply an off-the-shelf architecture, fine tuning it to our task, to show the advantages of knowledge transfer when working with limited data in the software domain. We focus on the classification of unified modeling language (UML) diagrams into class and sequence diagrams from a publicly-available dataset \cite{Hebig}. This dataset has been previously leveraged to demonstrate barriers that arise when applying deep architectures with vast parameter spaces.

\section{Transfer Learning}
Transfer learning is the process of taking a model trained for one task, where data is more readily available, and applying it to a new but similar task. Traditionally, given two separate tasks, we would have to obtain two distinct training sets and build models for each task. Unfortunately, large amounts of data in every domain are not always available, and in a lot of cases are not always needed if two tasks are similar enough. 

When considering how humans learn to do new tasks, they rarely have to start at the absolute beginning - tabula rasa - and typically are building off of similar previous experiences. If one tries to learn a new language, or how to play a new game, one draws on prior knowledge and adapts to complete the task at hand. This is core idea of transfer learning; to learn general features in one domain and apply those features to another, similar domain. In our case, we transfer general features learned when the VGG network has been trained on the ImageNet \cite{imagenet_cvpr09} dataset and fine tune it to the task of UML classification. We choose this classification task for our experiment for two reasons. First, the work in \cite{Ott2019} used this same data to demonstrate the inability of deep networks such as VGG-16 to learn features when training samples are limited, requiring custom architectures to be built. Second, UML is sufficiently dissimilar to other objects found in ImageNet that we can be confident that pre-trained models will not have already learned features directly applicable to the classification task.

When applying transfer learning, a decision must be made to determine how much will be borrowed from the original model. It is common practice to take an established architecture and freeze some amount of the original layers, while fine tuning the rest to the specific needs of a problem. As a result, only the unfrozen layers are trained - resulting in far fewer learnable parameters which decreases the size of the required labeled dataset for training. The amount frozen and fine-tuned is variable depending on the task at hand. We will explore two variations on the VGG-16 architecture, as well as a shallow CNN in this paper. In one VGG network, we fine tune all available weights and see poor accuracy when dealing with small training samples due to the large parameter space that must be learned. In the second, we freeze the majority of weights while fine tuning only the final layer and see accuracy near 90\% even at very low numbers of training samples. 

In general, when implementing transfer learning, we look in three areas for possible superiority over other networks, as outlined in \cite{Olivas}. First, we may find a higher starting accuracy, at the beginning of training, before the model has been refined further. Second, we could see a steeper or faster rate of improvement of accuracy as training continues. Finally, we look for a higher asymptote, or greater accuracy toward the end of training. In our results, we find that the frozen VGG network exhibits higher accuracy in all three of these areas over the pre-trained VGG and a shallow CNN.

\section{Data}
From the Lindholmen Dataset\cite{Hebig}, an initial corpus of 14,815 portable network graphics (PNG) images of UML diagrams is obtained. That is then reduced to 13,359 images when only active UML diagrams are considered. Of the active diagrams, there were 11,319 Class Diagrams and 2040 Sequence Diagrams. We resize all images to 250x250 pixels. This dataset was chosen for its small size and its relation to software repositories. The VGG-16 networks we include in our tests have been trained on the ImageNet dataset which includes over 1,000,000 natural images belonging to 1,000 categories. Although the natural images of ImageNet and UML diagrams exist in quite different domains, we still see improvement in classification when using knowledge transfer. 

\section{Methods and Experiments}

\begin{figure}[h]
  \centering
  \includegraphics[width=\linewidth]{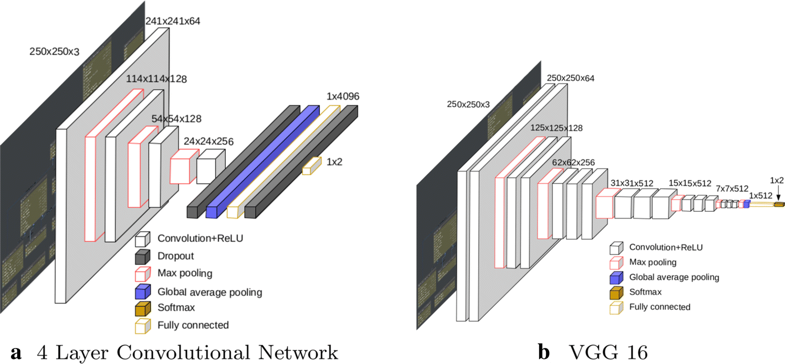}
  \caption{Visuals of the networks used a) The four convolutional layers, interspersed with max pooling for downsampling followed by dropout, max pooling, and fully connected layers for classification. b) Standard VGG network with sixteen convolutional layers. Image used with permission from \cite{Ott2019}.}
  
  \label{fig:networks}
\end{figure}

In our experiments, we compare three convolutional neural network (CNN) architectures on their classification ability of UML diagrams. First, we use a simple network with four convolutional layers, max pooling, dropout, and global average pooling layers followed by fully connected dense layers for classification. This network contains 2,260,000 trainable parameters. Two other networks explored are variations of the popular VGG network with sixteen convolutional layers modified to fit the size of our input data \cite{VGG}. The first VGG we test starts with the original weights and we then allow all 14,715,000 trainable parameters to be updated as we train for our task. Conversely, in the second VGG, we freeze the majority of layers, and then modify and train only the last layer. The four layer CNN and VGG-16 architectures are shown in Figure \ref{fig:networks}. All networks are implemented in Keras with a TensorFlow backend.

These three models were trained as binary classifiers to differentiate UML diagrams as either sequence or class diagrams. To show the advantages of transfer learning, we incrementally increase the available training data in two tests. We begin with 50 samples of each class and increase by increments of 250 to 1800 samples. A second test to show the accuracies at very low samples is performed beginning with 5 samples and increasing by increments of 5 to 50 samples. Upon incrementing the sample size, each network is reset to the same original weights.

Each model was trained for a minimum of 5 epochs and stopped when the accuracy had not improved after a patience of 5 epochs. We implemented 5 fold cross validation for robustness. 

The code and data to train all models, as well as the learned models themselves, are available publicly at: (removed for anonymity) We hope they, in turn, will be utilized for transfer learning in future deep learning applications on software data.

\section{Results}

\begin{figure}[h]
  \centering
  \includegraphics[width=\linewidth]{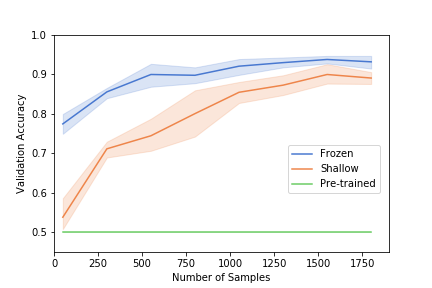}
  \caption{Displayed above is the number of training samples from each category, from 50 to 1800, vs the validation set accuracy achieved by each model. For robustness, 5 trials were run for all training samples tested. The color bands indicate the distribution of results from the 5 trials.}
  \label{fig:allsamples}
\end{figure}

Figure \ref{fig:allsamples} shows the test accuracy achieved by each network from 50 to 1800 samples of each class, or 100 to 3600 total images respectively. Both the frozen VGG and 4 layer CNN are eventually able to classify the given diagrams with about 90\% accuracy given a sufficient amount of samples. Although, we see a significant difference in the starting accuracies as well as faster convergence.

However, we are also interested in the best accuracy achievable with the least amount of data. The frozen VGG-16 is able to classify with an about 80\% accuracy after only 100 total training samples while the 4 layer CNN falls short at about 52\% accuracy. As can be expected, the VGG-16 that was left free to train the massive number of parameters within its network, also performs poorly, barely reaching 50\% accuracy for random guessing. The tiny amount of training data given to this network is, of course, nowhere near enough to train all 14 million parameters.

\begin{figure}[h]
  \centering
  \includegraphics[width=\linewidth]{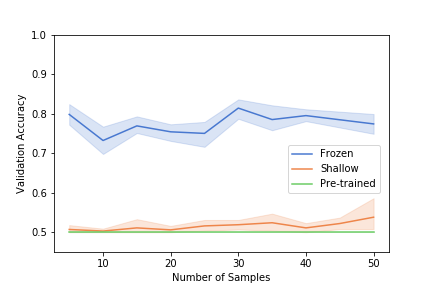}
  \caption{Shown above is the accuracy achieved by each network at the corresponding sample sizes, from 5 to 50 samples in each UML category. For robustness, 5 trials were run for all training samples tested. The color bands indicate the distribution of results from the 5 trials.}
  \label{fig:5to50}
\end{figure}

Figure \ref{fig:5to50} shows the training accuracy for all three networks when given 5 to 50 samples of each class, or 10 to 90 total images. We include this figure to demonstrate the superiority of the frozen VGG over both networks especially at very low samples. Even with only 10 total samples, the frozen VGG is able to classify the UML diagrams with an average 73\% accuracy, compared to an accuracy of only 50\% for both other networks.

\begin{figure}[h]
  \centering
  \includegraphics[width=\linewidth]{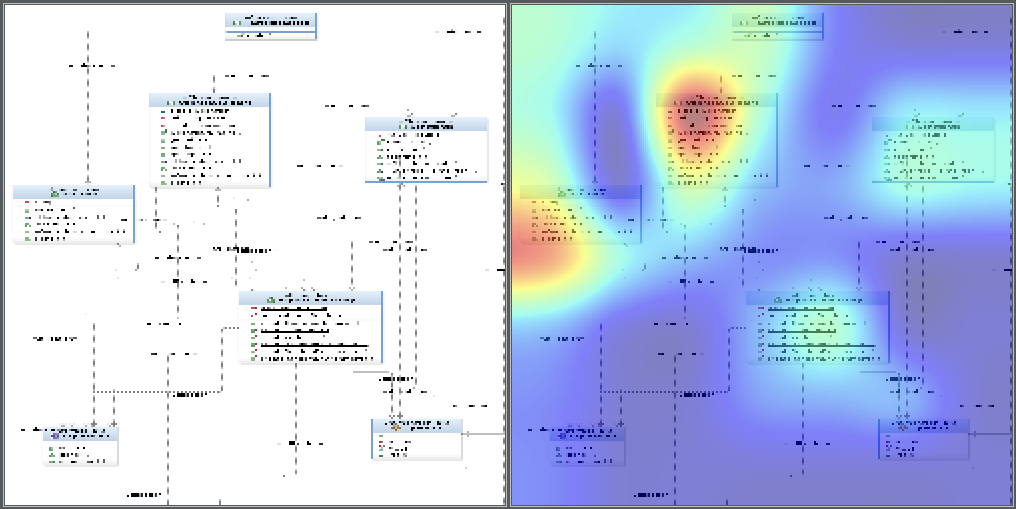}
  \caption{Class activation mapping prediction for a selected UML class diagram, original image on the left, heatmap indicating significant features on the right}
  \label{fig:class}
\end{figure}

\begin{figure}[h]
  \centering
  \includegraphics[width=\linewidth]{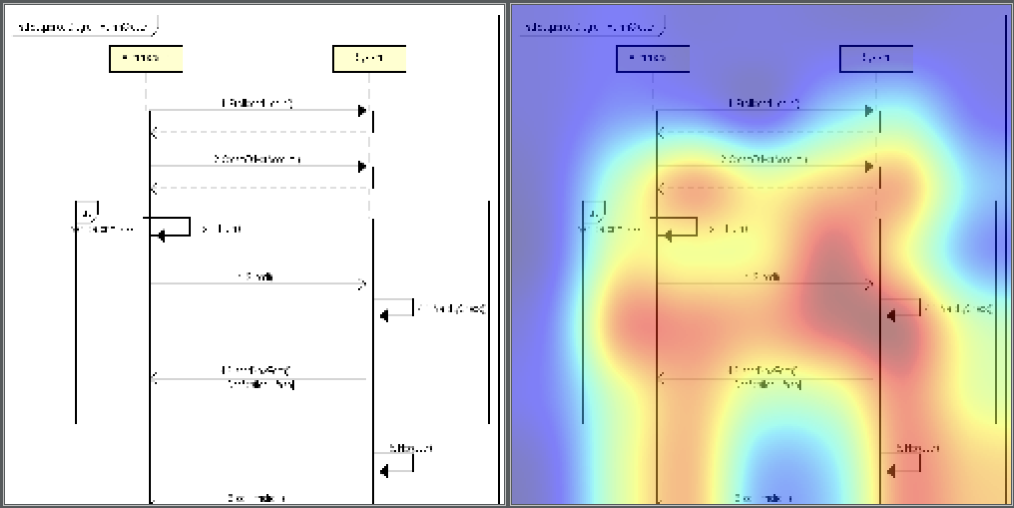}
  \caption{Class activation mapping prediction for a selected UML sequence diagram, original image on the left, heatmap indicating significant features on the right}
  \label{fig:sequence}
\end{figure}

Class activation mapping allows us to investigate further what parts of an image a convolutional network uses to make its prediction, as well as ensure those features make sense \cite{CAM}. Using the Keras Visualization Toolkit \cite{raghakotkerasvis}, we produced CAM results for one UML sequence diagram and one class diagram. CAM results are shown in figures \ref{fig:class} and \ref{fig:sequence} for the frozen VGG16 network trained on 1800 sample images from each class. CAM produces a heat map highlighting the regions most heavily weighted by the network. We are able to see clearly that the network learns features specific to sequence and class diagrams. Specifically, in class diagrams, the boxes containing class attributes and methods have been highlighted. Conversely, in sequence diagrams, the vertical lifelines are more significant.

\section{Related Work}
The classification of UML diagrams has been studied through a variety of machine learning techniques. Ho-Quang et al. \cite{HoQuang2014} proposed a logistic regression model using 19 of their 23 proposed features for classifying UML and non-UML class diagrams (CD). When trained on a corpus of 1300 images, their model achieved 96\% accuracy for UML-CD and 91\% of accuracy for non-UML CD. Years later, Ho-Quang et al. furthered their work to differentiate between diagrams that were hand-made as part of the forward-looking development process (FwCD), and diagrams that were  reverse  engineered  from  the source  code (RECD) \cite{HoQuang2018}. However instead of classifying the images directly, the authors extract various features and implement a random forest model to achieve 90\% accuracy in distinguishing the two types of class diagrams. In another study, using a corpus of 1300 UML and non-UML images, Hjaltason et al. trained a support vector machine (SVM) with an average classification accuracy of 92.05\% \cite{Hjaltason}. 
Moreno et al. conducted a similar study to classify web images as UML and non-UML class diagrams using a rule based approach. By extracting features from the images, in a corpus of 19000 web images, their algorithm reached an accuracy of 95\% \cite{Moreno}.

While we believe this is one of the first attempts to study the applicability of transfer learning to software engineering artifacts, transfer learning in general has been studied in many domains and aided in the development of powerful machine learning models. Shin et al. investigated the effectiveness of CNN architectures and transfer learning in detecting thoraco-abdominal lymph nodes and classifying interstitial lung disease \cite{shin}. The authors achieve state-of-the art performance and find transfer learning to be beneficial despite the natural images used to train ImageNet being significantly different from medical images. Another study applied transfer learning to four medical imaging applications in 3 specialties including radiology, cardiology, and gastroenterology \cite{Tajbakhsh}. Their experiments transferred weights from ImageNet layer-wise, using none, a few, or many layers and found that transferring a few layers improved performance compared to training from scratch. 

As stated previously, transfer learning in the space of software imagery was motivated by the work in \cite{Ott2019}. Here the authors showed definitively deep networks like VGG were unable to compete with smaller architectures when labeled data was sparse. A viable workaround was to create custom, shallower architectures that were compatible with available data volumes. The work presented here shows that off-the-shelf architectures can be used, but demand more efficient learning solutions - specifically the kinds produced via transfer learning. 

\section{Conclusion}

Transfer learning allows us to take, in effect, a shortcut in training deep architectures. In this paper, we extended previous work regarding the application of machine learning techniques for classification of UML images. Given limited data, it is nearly impossible to train a network with the depth and substantial number of parameters as in VGG. However, by transferring knowledge learned from one task to another task, we are able to tune off-the-shelf deep architectures and achieve high classification accuracy, rather than having to design new architectures with fewer layers and smaller parameter spaces to learn. Most importantly, the knowledge that forms the basis of the transfer learning needs no previous exposure to artifacts from the software domain, suggesting that transfer learning can be applied broadly to applications of deep learning within empirical software engineering.

Our experimental results have show training is positively effected by transfer learning even when the number of samples shown to the network is kept small. In contrast, even a smaller network with substantially fewer parameters is unable to learn as well. As a control, we also tested an off-the-shelf VGG and allowed the entire architecture containing over 14 million parameters to train. As expected, this network failed to improve beyond 50\% accuracy even when shown the maximum number of samples tested.

In addition to affirming the utility of utilizing transfer learning for mining software artifacts, our results suggest that as a research community we should be more proactive in curating machine learning models trained on software data, in addition to the software data itself. Such repositories of pre-trained models would allow empirical software engineering researchers to apply transfer learning to new applications using models already tuned using software data of various types. 

\printbibliography

\end{document}